\documentclass[11pt]{article}

\usepackage[margin=1in]{geometry}
\usepackage{amsmath, amssymb, amsfonts}
\usepackage{graphicx}
\usepackage{algorithm}
\usepackage{algorithmic}
\usepackage{subcaption}
\usepackage{textcomp}
\usepackage{xcolor}
\usepackage{float}

\title{Breaking Local-Minimum Traps in Spiking Neural Network-Based Solvers for CSPs via Parallel Tempering}
\author{
  Recep Bugra Uludag \\
  University of Minnesota - Twin Cities \\
  \texttt{uluda002@umn.edu}
  \and
  Ahmet Efe \\
  University of Minnesota - Twin Cities \\
  \texttt{efe00002@umn.edu}
  \and
  Ismail Akturk \\
  Ozyegin University \\
  \texttt{ismail.akturk@ozyegin.edu.tr}
  \and
  Ulya R. Karpuzcu \\
  University of Minnesota - Twin Cities \\
  \texttt{ukarpuzc@umn.edu}
}

\date{}
\begin{document}
\maketitle

\begin{abstract}
Spiking neural networks (SNNs) with stochastic neurons can solve constraint satisfaction problems (CSPs) by encoding constraints via connectivity and performing probabilistic search via spike dynamics. However, fixed-temperature stochastic dynamics often get trapped in local minima---near-satisfying configurations---a vulnerability that escalates with problem difficulty. To overcome this, we integrate parallel tempering (PT) into the neural sampling solver, running multiple parallel replicas at varying inverse temperatures. Replicas periodically exchange temperatures rather than network states, managing the trade-off between exploration and concentration around low-energy configurations while preserving asynchronous, spike-based computation. We evaluate this architecture against a parallel baseline of four independent, fixed-temperature solvers using equal computational resources across 1000 instances from the SATLIB uf20-91 benchmark. Parallel tempering improves success probability on 332 instances while worsening only 5. Crucially, these gains are concentrated on hard instances where independent solvers fail.
Violation trajectory analysis confirms the underlying mechanism: temperature exchanges allow replicas to traverse energy barriers unreachable by fixed-temperature dynamics, successfully escaping the narrow basins that constrain the baseline. To our knowledge, this represents the first integration of parallel tempering into an SNN-based CSP solver.
\end{abstract}

\section{Introduction}
Constraint satisfaction problems (CSPs) are mathematical problems in which values must be assigned to variables so that a set of constraints is satisfied~\cite{dechter2003CSP}. Many important real-world decision-making tasks can be formulated as CSPs or closely related constraint-based optimization problems, including digital circuit verification~\cite{gupta2006sat}, vehicle routing in logistics networks~\cite{shaw1998,toth2014}, and resource allocation in cloud and data-center infrastructures~\cite{cambazard2013}. Despite this broad applicability, CSPs are often computationally challenging. Many classes of CSPs are NP-complete, and practical search procedures frequently become trapped in local minima or near-satisfying configurations that hinder efficient discovery of valid solutions~\cite{hoos2004}. 

These challenges motivate the exploration of alternative search paradigms, particularly stochastic neural approaches that can support probabilistic exploration of complex solution spaces~\cite{pecevski2011}.
In these approaches, constraints are encoded directly in the network connectivity, and stochastic spike dynamics perform probabilistic search through the space of configurations. The neural sampling framework \cite{buesing_neural_2011} provides a principled foundation for this approach, establishing conditions under which stochastic spiking dynamics approximate sampling from an energy-based probability distribution defined by the network structure. Within this framework, solving a CSP is equivalent to searching for low-energy configurations in {the resulting} energy landscape, and the stochastic dynamics naturally favor lower-energy regions while retaining some probability of visiting higher-energy states. 

However, this stochastic search process remains vulnerable to local-minimum trapping. The network dynamics may spend substantial time in partially satisfying assignments that impede progress toward a valid solution. This limitation becomes increasingly severe as problem difficulty grows and is a major limitation on hard instances. Escaping such configurations requires more than simply increasing stochasticity, which can disrupt low-energy refinement; instead, it calls for a mechanism that expands exploration while preserving the ability to exploit promising regions of the search space. 

A natural way to address this challenge is to regulate the degree of stochastic exploration through a temperature parameter that controls the randomness of configuration transitions~\cite{Kirkpatrick_Optimization_1983}. Among the methods built on this principle, parallel tempering (PT), also known as replica exchange Monte Carlo, has proven effective for navigating rugged energy landscapes in classical optimization and sampling settings~\cite{Hukushima_Exchange_Monte_Carlo_1996}. PT runs multiple replicas of the system in parallel at different temperatures and periodically couples them through exchange operations. High-temperature replicas explore the state space broadly by more readily traversing high-energy regions, whereas low-temperature replicas focus on refining promising low-energy configurations. By allowing replicas to alternate between exploratory and exploitative regimes, PT enables the overall system to escape basins that would otherwise trap fixed-temperature dynamics.

\begin{figure}[t!]
\centering
\begin{subfigure}[b]{0.42\textwidth}
\centering
\includegraphics[width=\linewidth]{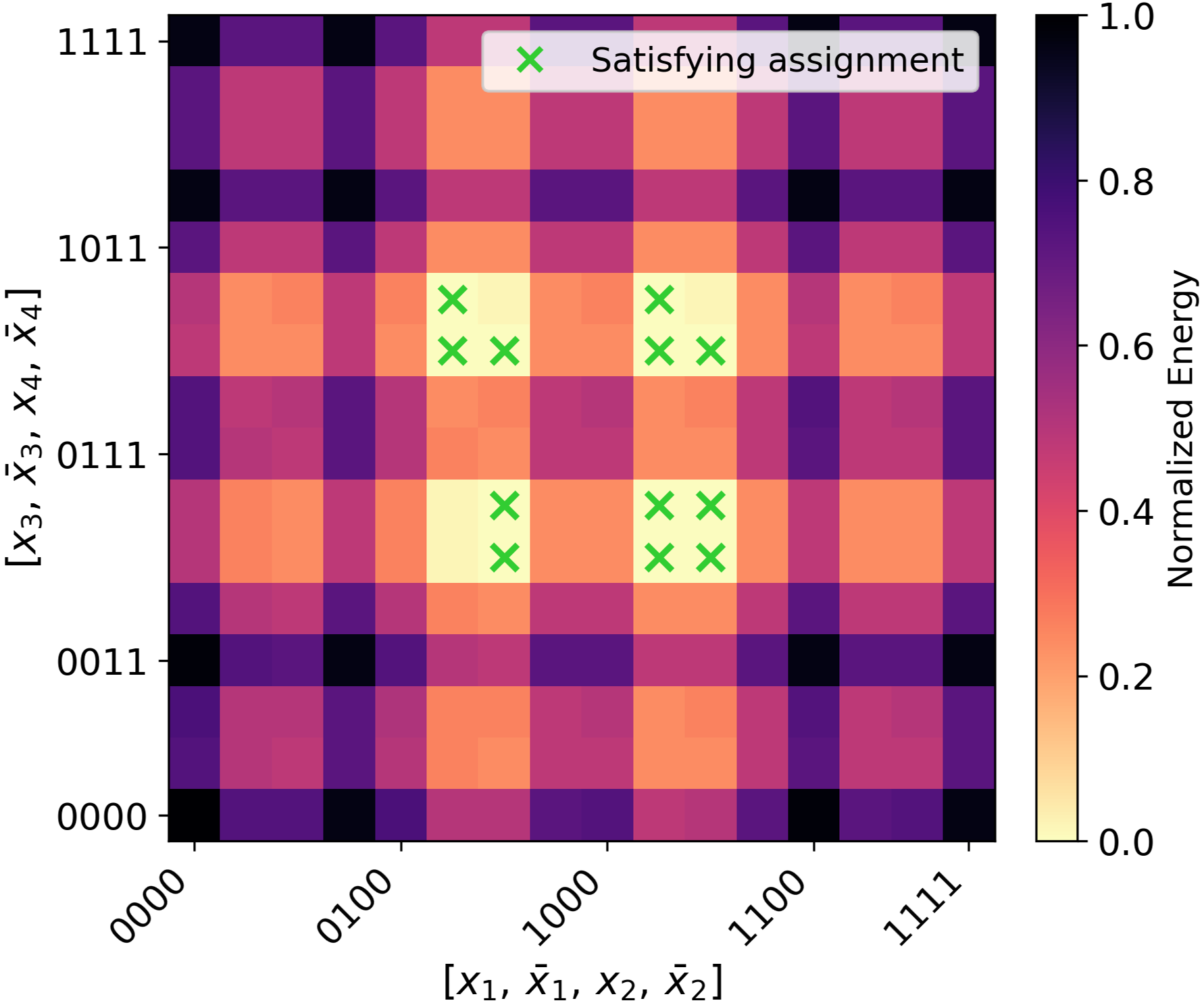}
\caption{Analytical energy landscape for the SAT formula
$(x_1 \vee x_2 \vee x_3) \land (\neg x_2 \vee \neg x_3 \vee \neg x_4)$.
Each cell represents a configuration of the binary variables. Cross marks (in green) mark satisfying assignments.}
\label{fig:energy_landscape}
\end{subfigure}
\hfill
\begin{subfigure}[b]{0.55\textwidth}
\centering
\includegraphics[width=\linewidth]{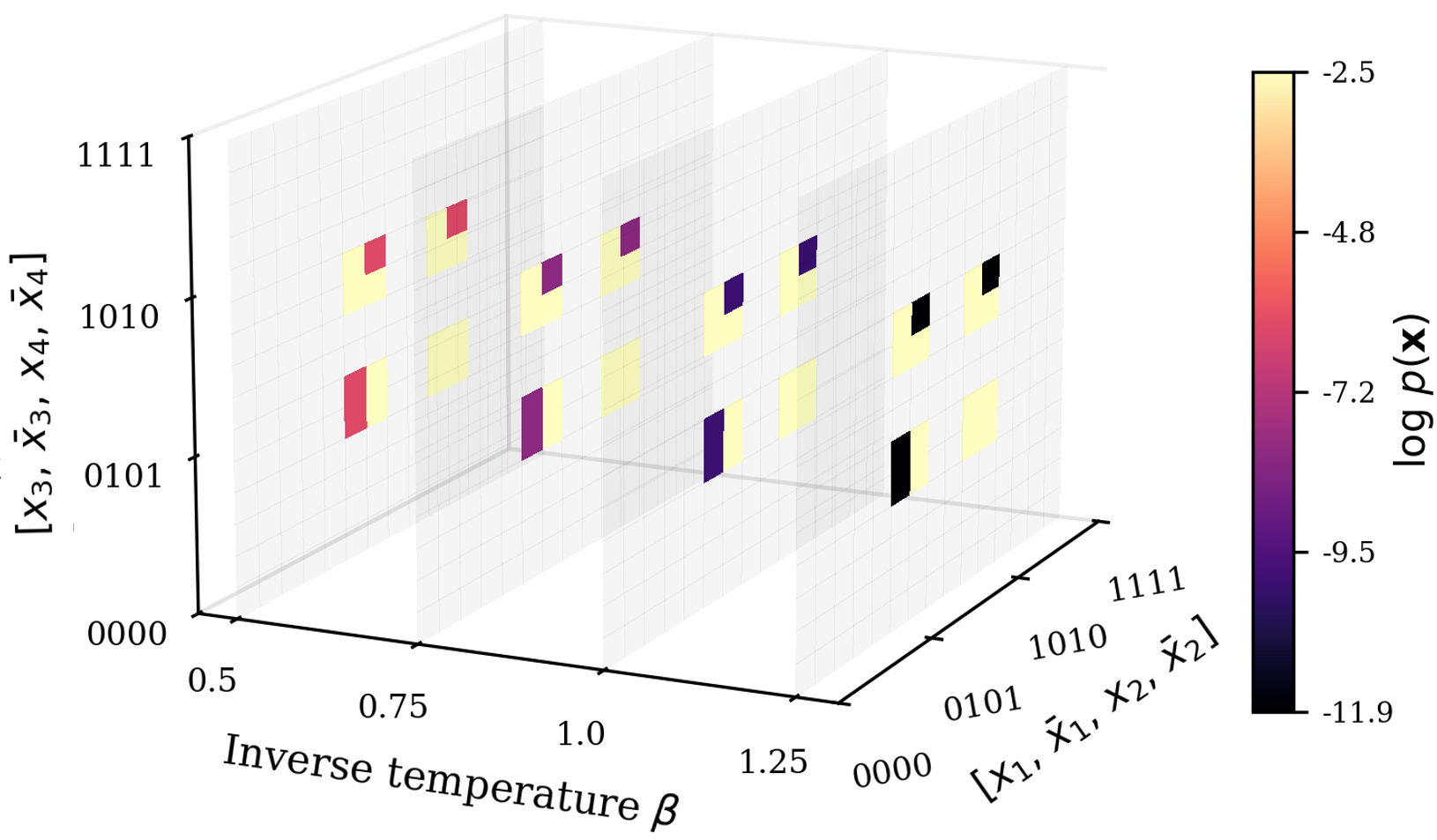}
\caption{Boltzmann probability distribution over the same configuration space at different inverse temperatures $\beta=1/T$. Higher $\beta$ concentrates probability mass around low-energy configurations, 
illustrating how temperature controls the progress of the stochastic search.
}
\label{fig:temperature_effect}
\end{subfigure}
\caption{Energy landscape and temperature-controlled sampling for a representative SAT problem.}
\label{fig:energy_temperature}
\end{figure}

In this paper, we integrate PT into an SNN-based solver for CSPs with stochastic neurons. We introduce temperature control into the neural sampling dynamics and run multiple solver replicas in parallel at different inverse temperatures, which regulate the balance between broad exploration and concentration around low-energy configurations. To respect the asynchronous spike-based dynamics of the network, replica coupling is implemented through temperature exchange rather than explicit state transfer. This design avoids moving internal dynamical variables between replicas while preserving the correct joint stationary distribution. To the best of our knowledge, this is the first integration of PT into an SNN-based CSP solver with stochastic neurons.

We evaluate the resulting method on the SATLIB benchmark suite~\cite{hoos_satlib_2000} and analyze its behavior at multiple levels. To ensure a fair comparison under equal computational resources, we compare PT against a parallel baseline consisting of four independent neural samplers running simultaneously at fixed temperature, thereby isolating the effect of inter-replica coupling from parallelism alone. A large-scale study over 1000 problem instances quantifies per-instance improvements in success probability and shows that the advantage of PT is concentrated on genuinely hard instances, where local-minimum trapping is the main obstacle to solution discovery. Instance-level cumulative analysis further shows that this benefit persists throughout runtime rather than arising only from early exploratory behavior. Finally, trajectory-level analysis of clause-violation dynamics reveals the mechanism directly: temperature exchange enables replicas to reach energy regions that independent fixed-temperature solvers fail to access. Together, these results show that PT improves the robustness of neural sampling-based CSP solving precisely in the regime where independent parallelism is insufficient.\label{Introduction}

\section{Background}

\begin{figure}[t!]
\centering
\begin{subfigure}[t]{0.47\linewidth}
\centering
\includegraphics[width=\linewidth]{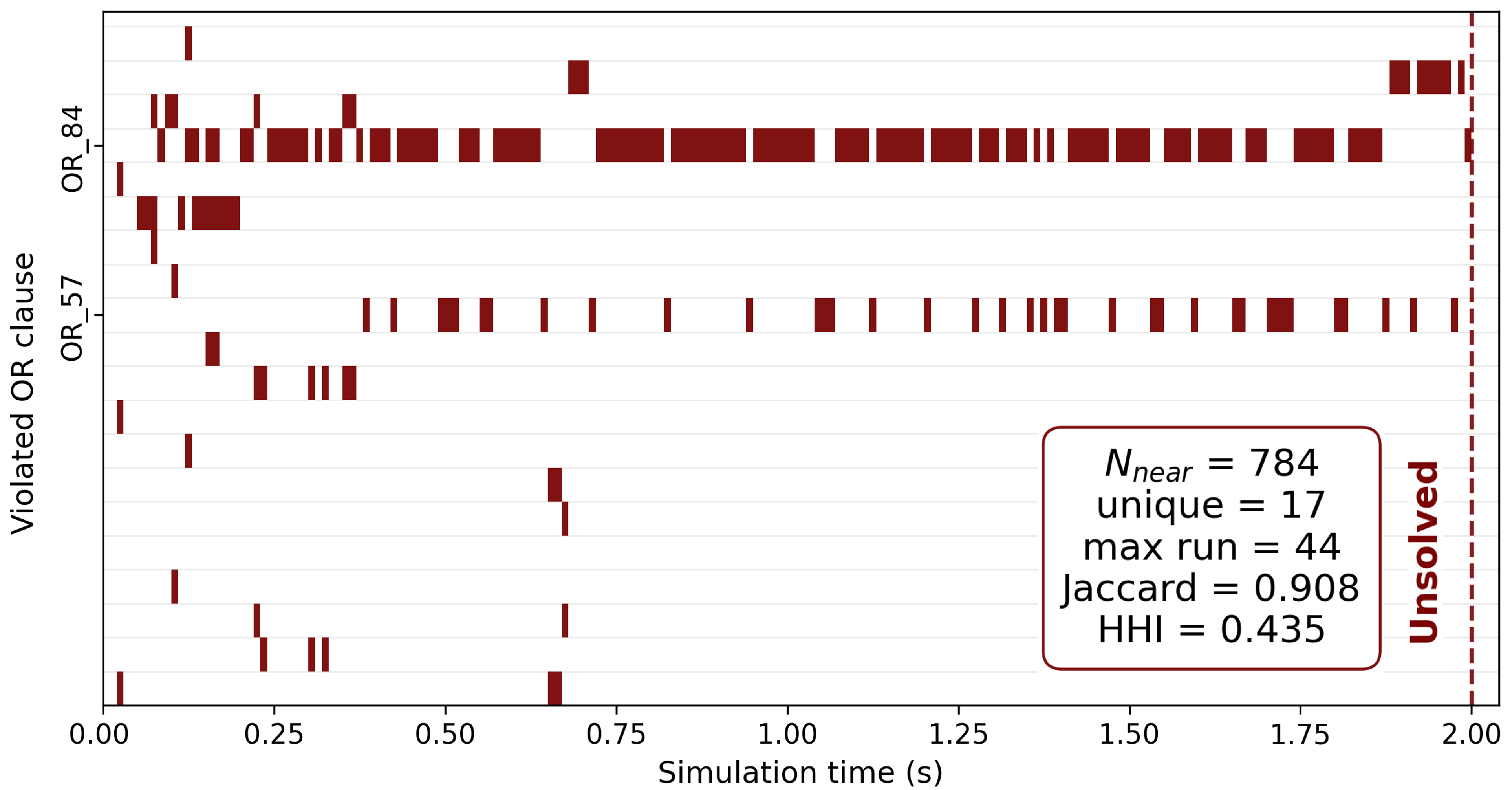}
\end{subfigure}
\hfill
\begin{subfigure}[t]{0.47\linewidth}
\centering
\includegraphics[width=\linewidth]{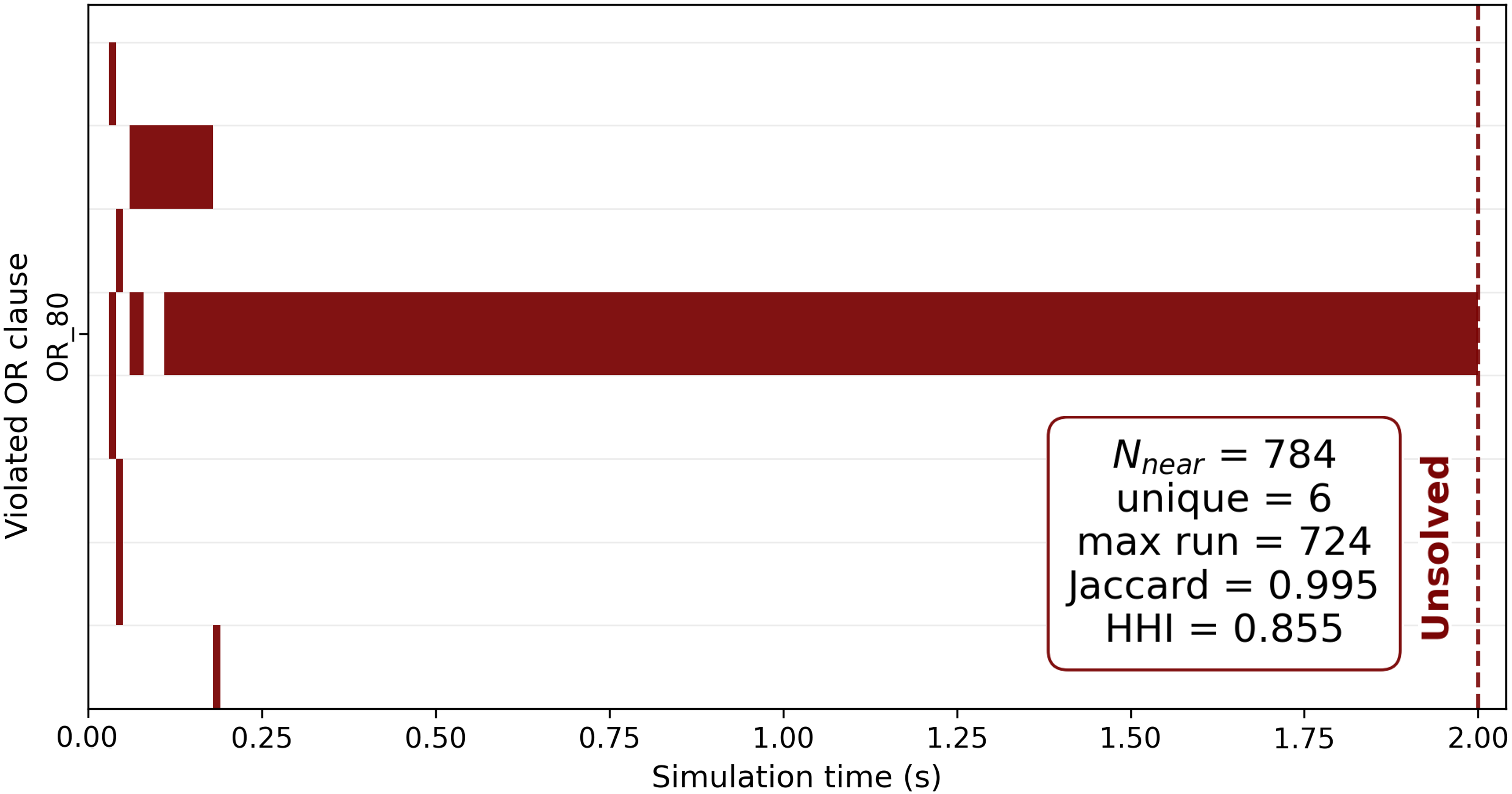}
\end{subfigure}
\caption{
Neural CSP solver trapped in local minima in solving a representative SAT problem.
\textbf{Left:} the solver cycles among a 
small number of violation signatures, producing a bouncing pattern between 
nearby near-solution configurations (HHI\,=\,0.435, max run\,=\,44). 
\textbf{Right:} the solver becomes trapped in a single dominant violation 
configuration for nearly the entire runtime (HHI\,=\,0.855, max run\,=\,724). 
Such local minima traps prevent the dynamics from 
reaching a satisfying assignment.
}
\label{fig:local_minima_example}
\end{figure}


The proposed method builds on three key ideas: i) the representation of CSPs as energy landscapes, ii) the use of SNNs with stochastic neurons for probabilistic search, and iii) the observation that fixed-temperature neural dynamics can become trapped in local minima. This section reviews each of these components in turn and thereby establishes the basis for the parallel tempering framework introduced later.

\subsection{From CSPs to Energy Landscapes}

Constraint satisfaction problems (CSPs) require assigning values to variables from discrete domains such that all constraints are simultaneously satisfied. This formulation captures a broad class of combinatorial search problems, for which the number of possible assignments typically grows exponentially with problem size. As a result, finding a satisfying assignment, or even determining whether one exists, can be computationally hard.

A useful way to study such problems is through an energy-based representation that assigns each configuration $x$  an energy value $E(x)$. In this view, constraint violations contribute energy penalties, so configurations satisfying more constraints lie in lower-energy regions of the landscape. Solving the CSP is therefore equivalent to searching for a zero-violation configuration, or more generally, for a global minimum of the corresponding energy function.

In the context of SNN-based CSP solving, this representation is more than a convenient abstraction. By assigning a scalar energy to each configuration, it provides a graded objective that distinguishes among partially satisfying states while ensuring that satisfying assignments correspond to global minima. This makes the problem well suited to stochastic neural dynamics, which can be interpreted as probabilistic exploration of the resulting energy landscape rather than purely symbolic constraint checking.

This representation also naturally supports stochastic search through a probabilistic distribution over configurations. Rather than traversing the landscape deterministically, the solver explores configurations by sampling from a Boltzmann distribution,

\begin{equation}
    p(x) \propto \exp\!\left(-\beta E(x)\right)
\end{equation}

\noindent where $\beta$ denotes the inverse temperature (i.e., $1/T$). Small values of $\beta$ reduce the influence of energy differences and promote broad exploration, whereas large values concentrate probability mass around low-energy configurations, encouraging refinement toward promising solutions.

Figure~\ref{fig:energy_temperature} illustrates this relationship for a representative SAT instance. Figure~\ref{fig:energy_landscape} shows the corresponding energy landscape, while Figure~\ref{fig:temperature_effect} shows how the probability distribution becomes increasingly concentrated around low-energy configurations as the inverse temperature \(\beta\) increases. These figures highlight the central trade-off between broad exploration of the state space and concentration around favorable configurations.

\subsection{Neural Sampling for CSP Solving}

Building on the energy-based formulation above, neural sampling approaches treat CSP solving as probabilistic exploration of the configuration space. 

The neural sampling framework for CSP solving was introduced by Jonke et al.~\cite{jonke_solving_2016}, who showed that stochastic SNN dynamics can be structured to sample from Boltzmann distributions over constraint-encoded energy landscapes. In neural sampling, the stochastic search process is implemented by a recurrent network of spiking neurons. Subsequent implementations on neuromorphic platforms further demonstrated the viability of this approach~\cite{fonseca_guerra_using_2017,binas_spiking_2016,Davies_Advancing_2021}. The key idea is to interpret network activity over a short temporal window as a binary configuration and to design the network so that its stochastic dynamics sample from a target distribution over those configurations. Let $x = (x_1, x_2, \dots, x_N)$ denote the network configuration, where

\begin{equation}
    x_i(t) =
\begin{cases}
1 & \text{if neuron } i \text{ spikes in } (t-\tau,\,t] \\
0 & \text{otherwise,}
\end{cases}
\end{equation}

and $\tau$ is the duration of the rectangular postsynaptic potential, taken equal to the refractory period.

The membrane potential of neuron $i$ is given by

\begin{equation}
    u_i(t) = b_i + \sum_j w_{ij}\, x_j(t)
\end{equation}

\noindent where $b_i$ is the bias of neuron $i$ and $w_{ij} \in \mathbb{R}$ is the synaptic weight from neuron $j$ to neuron $i$. Spikes are generated stochastically at instantaneous rate

\begin{equation}
    \lambda_i(x) = \frac{1}{\tau}\exp\bigl(u_i(x)\bigr)
\end{equation}

Under the Neural Computability Condition~\cite{buesing_neural_2011}, these spike-driven transitions define a continuous-time Markov process whose stationary distribution is Boltzmann:

\begin{equation}
    p(x) \propto \exp(-E(x))
\end{equation}

For pairwise interactions, the corresponding network energy is

\begin{equation}
    E_N(x) = -\sum_i b_i x_i - \frac{1}{2}\sum_{i,j} w_{ij}\, x_i x_j
\end{equation}

This baseline form captures only pairwise dependencies. To represent structured CSP constraints, auxiliary circuit motifs can be added so that they contribute extra energy terms~\cite{jonke_solving_2016}:

\begin{equation}
    E(x) = E_N(x) + \sum_k U_k(x)
\end{equation}

where $E_N(x)$ denotes the baseline energy induced by pairwise interactions among variable neurons, while each $U_k(x)$ is an additional energy term contributed by an auxiliary motif that encodes a higher-order constraint. Examples include Winner-Take-All motifs, which enforce exclusivity among the possible values of a variable, and OR motifs, which encode logical clause structure in SAT instances~\cite{jonke_solving_2016}.

In this way, the network connectivity defines an energy landscape over candidate assignments, and the stochastic spike dynamics realize a sampling-based search over that landscape. The solver does not follow a fixed deterministic trajectory; instead, it continually moves among configurations, favoring lower-energy regions while retaining a nonzero probability of visiting higher-energy states. This probabilistic search mechanism is what makes neural sampling attractive for CSP solving, but it also gives rise to the local-minimum trapping behavior discussed next.

\subsection{Local Minima in Neural CSP Solvers}

Although stochastic neural sampling can escape shallow energy barriers, fixed-temperature dynamics may still become trapped in local minima corresponding to near-satisfying assignments. Figure~\ref{fig:local_minima_example} illustrates this behavior for a representative SATLIB instance (uf20-0192) in which pronounced trapping is observed within a 2 s simulation budget. Each row corresponds to an OR-clause motif, denoted OR$_k$, representing the $k$-th clause of the SAT formula.

Here, we focus on configurations with at most three violated clauses, which represent near-solution regions of the search space. Because SAT is a decision problem, these states are not approximate solutions in an optimization sense; rather, they are trapping configurations from which the dynamics must escape in order to reach a satisfying assignment. Each column in the visualization corresponds to one such configuration, and each row indicates whether a particular OR clause in the SAT formula is violated.

To characterize trapping behavior, we track the sequence of violation signatures, defined as binary vectors indicating which clauses are violated at each near-solution visit. For each run,$N_{\text{near}}$ denotes the total number of time steps spent in near-solution configurations, while unique counts the number of distinct violation signatures observed. Max run records the longest consecutive streak of identical signatures, capturing the persistence of a single trapping configuration. We also compute the average Jaccard index between consecutive signatures, which measures how similar successive near-solution states are, and the Herfindahl--Hirschman Index (HHI),

\begin{equation}
    \mathrm{HHI} = \sum_k p_k^2,
\end{equation}

\noindent where $p_k$ is the observed frequency of the $k$-th unique violation signature. An HHI of 1 indicates complete concentration on a single signature, whereas smaller values indicate more distributed exploration across near-solution states.

Figure~\ref{fig:local_minima_example} highlights two characteristic trapping behaviors. In the first case (left), the solver cycles among a small set of near-solution configurations. The violation pattern is dominated by OR$_{84}$ throughout most of the run, with only brief excursions to other signatures, indicating confinement to a narrow basin with limited variability (unique: 17, max run: 44, $\mathrm{HHI}=0.435$). In the second case (right), the solver becomes effectively trapped in a single dominant configuration early in the run. OR$_{80}$ remains the only violated clause for nearly the entire simulation, with almost no successful escape attempts (unique: 6, max run: 724, $\mathrm{HHI}=0.855$).

These traces show that failure is not simply the result of random wandering through a large search space. Instead, the solver enters near-solution basins that remain dynamically stable under fixed-temperature sampling. This exposes a fundamental limitation of single-temperature neural sampling: increasing stochasticity can improve escape from such basins, but doing so also weakens concentration around promising low-energy configurations. A more effective approach is therefore needed to combine broad exploration with reliable refinement.

Temperature provides a natural mechanism for controlling this trade-off. Higher temperatures reduce the influence of energy differences and promote exploration, whereas lower temperatures concentrate probability mass around favorable configurations and support refinement. This observation motivates the use of multiple replicas evolving simultaneously at different temperatures, so that exploratory and exploitative search modes can coexist within the same overall system. Parallel tempering provides exactly this capability, and the next section describes how we integrate it into the neural sampling framework.
\label{Background}

\section{Breaking Local-Minimum Traps via Parallel Tempering}
The analysis above motivates a search mechanism that can combine two competing requirements: broad exploration to escape local minima and focused refinement within promising low-energy regions. In this section, we show how these roles can be separated and coordinated through temperature. We first introduce temperature control into the neural sampling dynamics, thereby making the exploration/refinement trade-off explicit. We then use this temperature-controlled formulation to construct a parallel tempering method in which multiple replicas search simultaneously at different inverse temperatures.

\subsection{Temperature Control in Neural Sampling}

To introduce explicit control over the exploration-refinement trade-off, we incorporate an inverse temperature parameter $\beta = 1/T$ into the neural sampling dynamics. Specifically, $\beta$ rescales the membrane potential in the instantaneous firing rate of each neuron:

\begin{equation}
    \lambda_i(x) = \frac{1}{\tau} \exp\!\left(\beta\, u_i(x)\right).    
\end{equation}

This modification directly induces temperature dependence in the distribution sampled by the network. Under the Neural Computability Condition~\cite{buesing_neural_2011}, the membrane potential $u_i(x)$ equals the conditional log-odds of neuron $i$ being active given the current network state. Rescaling this quantity by $\beta$ is therefore equivalent to replacing the original energy function $E(x)$ with $\beta E(x)$ in the stationary distribution, yielding

\begin{equation}
    p_\beta(x) \propto \exp\!\left(-\beta E(x)\right).  
\end{equation}

The effect of $\beta$ is intuitive. Small values of $\beta$ reduce the influence of energy differences, flatten the effective landscape, and promote broad exploration across configurations. Large values sharpen the landscape, making low-energy states more strongly preferred and concentrating the dynamics around promising configurations. In this way, temperature provides a direct mechanism for controlling stochastic search behavior without modifying the network architecture or the underlying constraint encoding.

\subsection{Parallel Tempering}

PT addresses the exploration/refinement trade-off by running multiple replicas of the same SNN-based solver simultaneously at different inverse temperatures. All replicas share the same network architecture, synaptic weights, biases, and constraint encoding; they differ only in their assigned inverse temperature $\beta_r$. As a result, replica $r$ independently samples from $\exp\!\left(-\beta_r E(x)\right)$. In this work, we use an inverse-temperature ladder ranging from colder to hotter replicas, enabling a balance between low-energy refinement and broad exploration.

To couple these complementary search regimes, exchange attempts are performed every $\Delta t_{\mathrm{swap}} = 0.05$\,s of simulation time. At each exchange checkpoint, swaps are proposed between neighboring replicas in the temperature ladder using the standard Metropolis acceptance rule~\cite{hastings1970}. Let $(x_r,\beta_r)$ and $(x_s,\beta_s)$ denote the configurations and inverse temperatures of two adjacent replicas. The proposed exchange is accepted with probability

\begin{equation}
P_{\mathrm{swap}}
=
\min\!\left(
    1,\;
    \exp\!\left(
        (\beta_r - \beta_s)
        \left(E(x_r) - E(x_s)\right)
    \right)
\right).    
\end{equation}

This acceptance criteria ensures that the correct joint distribution over replica configurations and temperatures remains invariant under the exchange dynamics.

A key design choice in our setting is how exchanges are implemented. In conventional PT, one may view the algorithm as exchanging configurations between temperatures. However, the SNN solver evolves through asynchronous spike events in continuous time, so its instantaneous state is not just a static binary assignment, but also includes ongoing internal dynamical variables associated with spike timing and refractory activity. Explicitly transferring full network states between replicas would therefore require moving these internal dynamical variables across processes, introducing substantial implementation complexity and potentially disrupting the ongoing spiking dynamics. To avoid this difficulty, we implement replica coupling through temperature exchange: the inverse temperature parameters $\beta_r$ and $\beta_s$ are swapped, while each replica's internal neural state is left untouched. This is not only a practical simplification. Hukushima and Nemoto~\cite{Hukushima_Exchange_Monte_Carlo_1996} showed that configuration exchange and temperature exchange are statistically equivalent in PT, in the sense that they yield identical canonical expectation values for observables. In the present spiking-network setting, this equivalence allows us to preserve the correct joint stationary distribution while respecting the asynchronous nature of the neural dynamics.

Finally, exchange proposals follow an odd-even schedule, in which neighboring pairs are considered in two alternating rounds. This ensures that every adjacent temperature pair is periodically evaluated while avoiding conflicting simultaneous proposals, which would otherwise arise in ladders with more than two replicas.\label{Methodology}

\section{Evaluation}
\begin{figure}[t!]
    \centering
    \includegraphics[width=0.5\linewidth]{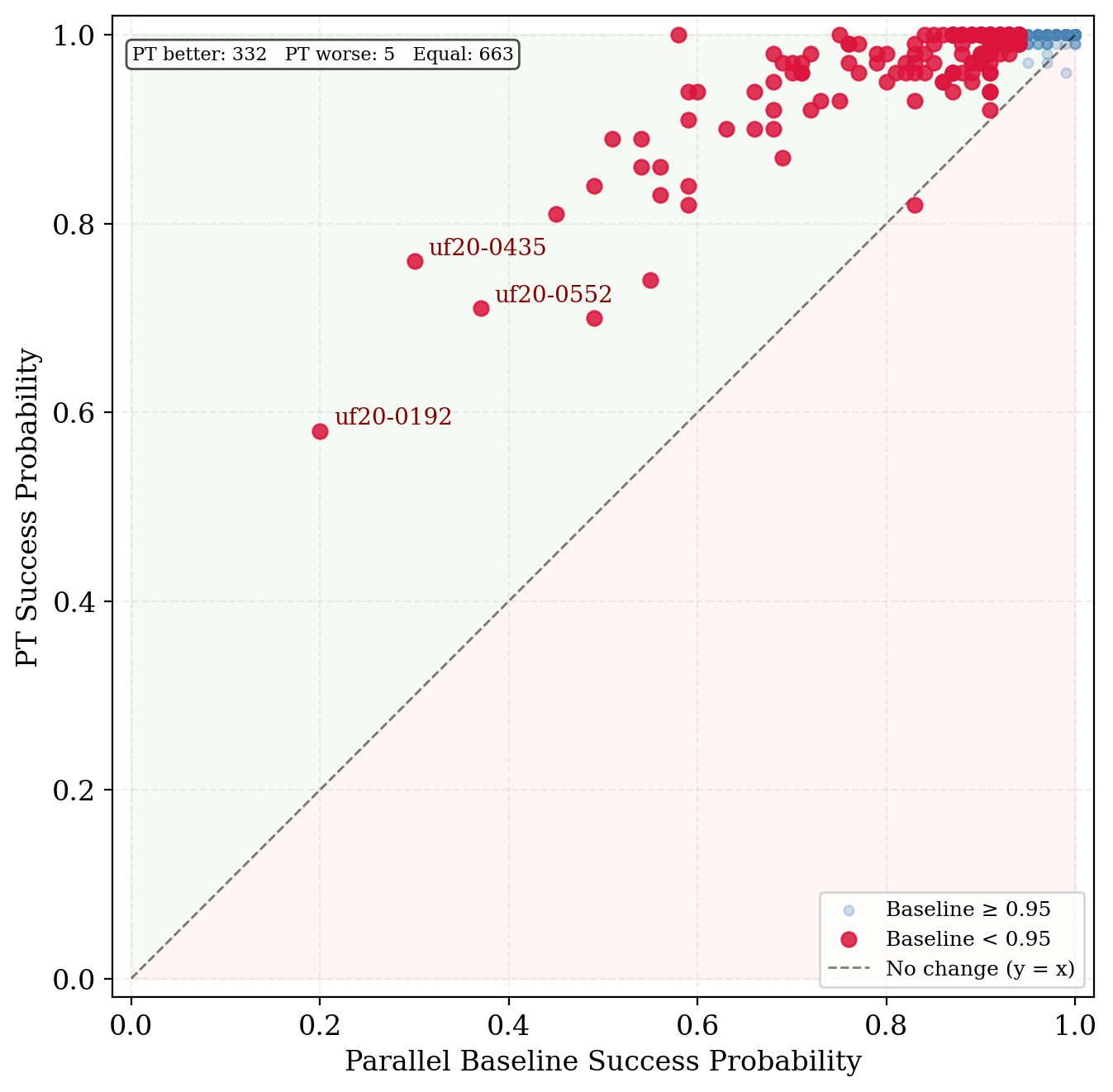}
    \caption{Per-instance success probability: PT versus parallel baseline across all $1000$ SATLIB \texttt{uf20-91} instances. Blue points indicate instances where the baseline achieves $\geq 0.95$ success probability; red points indicate harder instances. Points above the diagonal indicate PT improvement. The asymmetry (332 improved, 5 worsened) demonstrates that temperature exchange consistently benefits instances where independent parallel solvers are insufficient.}
    \label{fig:scatter}
\end{figure}


We evaluate the proposed PT-based neural sampler along three complementary dimensions. First, we measure per-instance success probability across the full benchmark suite to determine whether PT improves solver effectiveness overall. Second, we focus on the subset of hard instances where the parallel baseline fails more frequently, in order to assess whether PT's advantage is concentrated in the regime where local-minimum trapping is most severe. Third, we analyze runtime violation dynamics on representative hard instances to identify the mechanism by which PT alters the search process.

\subsection{Evaluation Setup}

To evaluate the proposed approach, we consider Boolean satisfiability (SAT), one of the canonical NP-complete constraint satisfaction problems. SAT has been extensively studied in both classical and neuromorphic optimization literature and serves as a standard benchmark for evaluating stochastic search algorithms.

We perform experiments on instances from the standard SATLIB benchmark suite \cite{hoos_satlib_2000}. SATLIB provides widely used benchmark problems designed to test the performance of SAT solvers near the phase transition region, where problem instances are known to be computationally challenging. In particular, we evaluate our solver on the \texttt{uf20-91} dataset, which contains randomly generated 3-SAT instances with 20 variables and 91 clauses. 

All experiments were run on a custom event-based simulator written in
Python 3.6.8. The simulator follows a similar methodology to NEVESIM~\cite{pecevski_nevesim_2014}, and maintains a central event heap and repeatedly
processes the earliest event, either distributing an outgoing spike or
releasing a refractory neuron; to reduce heap overhead, spike events use
version-based invalidation with periodic cleanup. 
Simulations were run on a
single node with an AMD EPYC 7763 (128 cores, 512\,GB DDR4) under
Linux 4.18.0. The Metropolis criterion is evaluated in Python using the
number of unsatisfied constraints as an energy proxy, read from each
replica's most recently recorded configuration. A replica-exchange step
therefore consists of a constant number of such reads followed by the
Metropolis test, which is negligible compared to the per-event neural
updates. This exchange logic is external to the neural dynamics and does
not advance the simulated clock, so it does not affect the reported timing
results.

For each instance, we run 100 independent trials under a fixed simulation time budget. A trial is considered successful if any replica reaches a satisfying assignment within the simulation time budget of $2$s.To ensure a fair comparison under equal computational resources, the baseline consists of 4 independent neural samplers running simultaneously at $\beta=1$ with no temperature exchange, where a trial is considered successful if any of the 4 solvers finds a satisfying assignment. PT likewise runs 4 replicas at temperatures $\beta \in \{1.25, 1.0, 0.75, 0.5\}$ with periodic replica exchange. This design ensures that any observed advantage of PT reflects the benefit of inter-replica temperature exchange rather than parallelism alone.

As the underlying neural CSP solver, we use the neural sampling architecture of Jonke et al.~\cite{jonke_solving_2016}. We adopt the same network structure, neuron model, and constraint-encoding motifs, including OR-circuit weights $W_{\mathrm{OR1}}=7.5$ and principal-neuron bias $b_{\mathrm{WTA}}=0$. Our implementation preserves the original network dynamics and augments them only with the PT mechanism introduced in Section~\ref{Methodology}.

\subsection{Benchmark-Level Performance}

We compare per-instance success probability for PT and the parallel baseline across all 1000 benchmark instances. Figure~\ref{fig:scatter} summarizes this comparison. The majority of instances, 856 out of 1000, are solved reliably by both methods at rates at or near 1.0, forming a dense cluster in the upper-right corner of the plot. These instances correspond to the easy regime, in which both methods already achieve near-ceiling performance and temperature exchange provides little additional benefit.

Across all 1000 instances, PT improves success probability in 332 cases while worsening only 5, a strongly asymmetric outcome that is unlikely to be explained by random variation alone. The remaining difference is concentrated in the hard regime, and the magnitude of improvement grows as baseline success probability falls, so the largest gains occur precisely on the instances that are hardest for the fixed-temperature parallel baseline.

To verify that these improvements are not artifacts of the finite trial
count, we quantified per-instance uncertainty and tested each instance
individually. We first computed Wilson score intervals on the per-instance
success rates. On easy instances (baseline rate $\geq 0.95$) the interval
half-width is at most $\pm 0.02$, so sampling noise is negligible and there
is no room for a detectable effect. On the 144 hard instances
(baseline $< 0.95$) the half-width reaches $\pm 0.10$, setting the scale of
noise relevant to the comparison. We then applied a two-sided Fisher's
exact test to each instance's success counts at $n=100$, with
Benjamini--Hochberg correction ($\alpha = 0.05$) applied across all tested
instances. Of the 144 hard instances, 89 (62\%) show a statistically
significant improvement under PT, while the single hard instance on which
PT is empirically worse does not reach significance. The significant cases
lie entirely in the hard regime, exactly where the Wilson intervals leave
room for a real effect.

Per-instance tests are necessarily underpowered at $n=100$, but the
aggregate direction is highly consistent: of the 144 hard instances, 143
favor PT and only 1 favors the baseline, and that single case lies within
its noise margin. A directional split of this magnitude is not plausibly
produced by chance, indicating that PT's advantage is systematic rather
than incidental.

We also compared mean time-to-solution on instances that both methods solve reliably. The resulting distributions overlap substantially, indicating that the two methods derive similar speed benefits from the parallelism they share. The main distinction is therefore not raw speed on easy instances, but coverage of hard instances that the baseline fails to solve consistently.

\subsection{Isolating the Role of Temperature Exchange}

The parallel baseline shares PT's replica count but holds all replicas at
$\beta = 1$, so it controls for parallelism but not for temperature
diversity. A natural concern is therefore whether PT's advantage stems from
inter-replica exchange specifically, or merely from running replicas at
different temperatures. To separate these two effects, we add a
diversity-only control: four replicas at the same varied temperatures as PT
($\beta \in \{1.25, 1.0, 0.75, 0.5\}$) but with no exchange between them.
Comparing PT against this control isolates the contribution of the exchange
mechanism itself, since the two configurations differ only in whether
replicas are allowed to swap temperatures. We ran this control on the 144 hard instances, where any exchange effect is most likely to be observable.

Across these 144 instances, PT outperforms the diversity-only control on
100, the control outperforms PT on 18, and the two are tied at a success
probability of 1.0 on the remaining 26. The per-instance gaps are small
enough that no single instance reaches Fisher significance, as expected:
the difference between PT and diversity-only is inherently narrower than
that between PT and the fixed-temperature baseline. The directional split,
however, is decisive: among the 118 instances that are not tied, a two-sided
sign test rejects the hypothesis of no exchange effect at the directional split,
however, is decisive: among the 118 instances that are not tied, a two-sided
sign test rejects the hypothesis of no exchange effect at $p < 10^{-10}$.
Temperature diversity alone therefore closes part of the gap to PT, but
inter-replica exchange contributes a further, directionally consistent
improvement that diversity alone does not account for.

\subsection{Focused Analysis on Hard Instances}

\begin{figure}[t!]
\centering
\includegraphics[width=.6\linewidth]{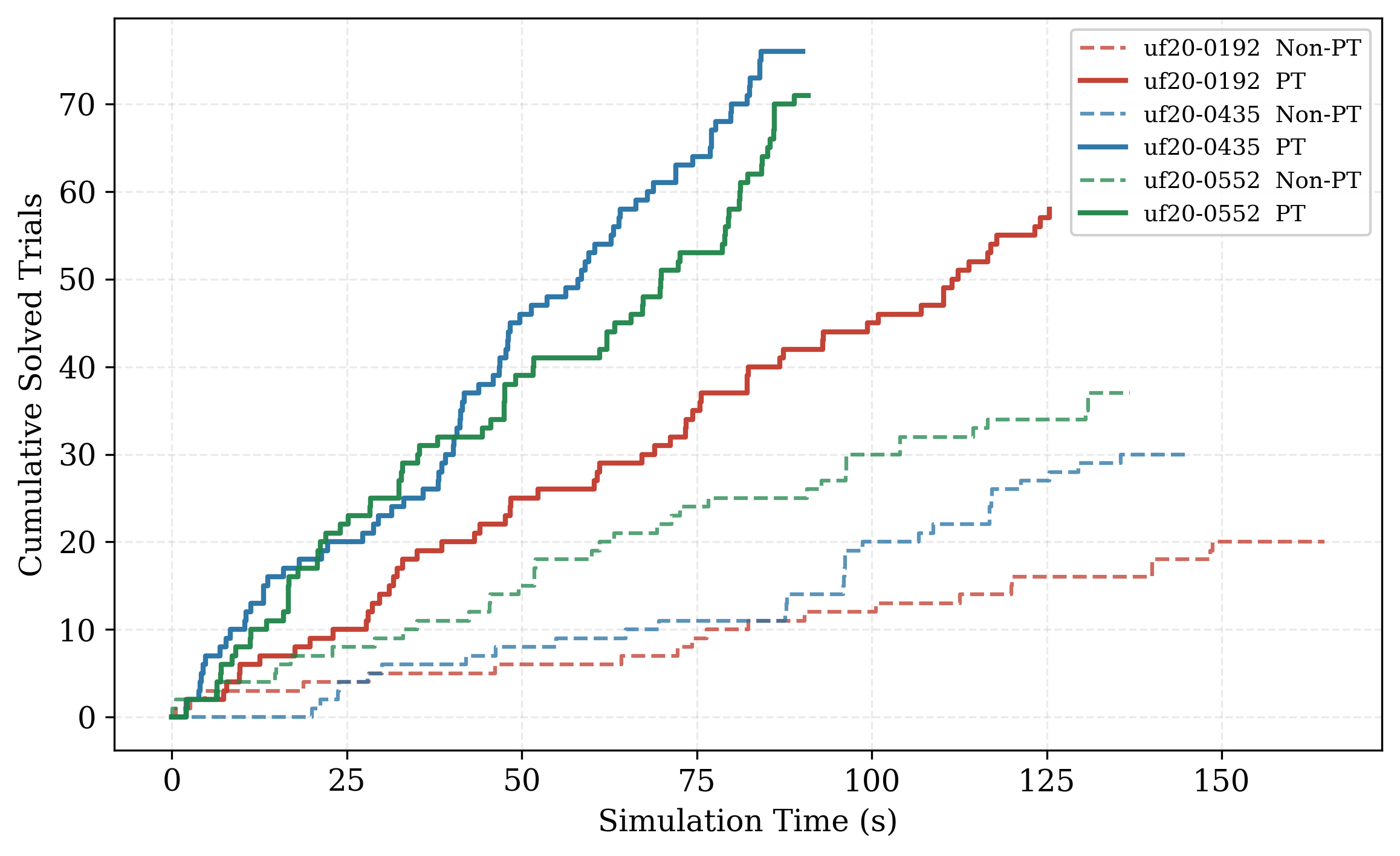}
\caption{Cumulative solved trials over time for the three hardest benchmark instances. Solid lines denote PT; dashed lines denote the parallel baseline. PT accumulates solutions more rapidly and consistently over time, while the parallel baseline exhibits extended stalling periods where all independent solvers are simultaneously trapped in local minima.
}
\label{fig:cumulative}
\end{figure}

Although Figure~\ref{fig:scatter} establishes that PT is most beneficial in the hard regime, it does not show how this advantage develops over time. To examine the temporal behavior more closely, Figure~\ref{fig:cumulative} reports cumulative success counts for the three hardest instances, as determined by baseline performance.


Across all three instances, the parallel baseline accumulates successful trials slowly and irregularly, with extended flat regions indicating periods during which all four independent samplers remain trapped and make no progress. PT, in contrast, accumulates successful trials at a substantially higher and more consistent rate, with the solid curves remaining well above their dashed counterparts from early in the runtime onward. This persistent separation shows that the benefit of PT is not the result of a single fortunate escape event. Rather, it reflects a sustained improvement in search effectiveness over the full runtime, consistent with repeated relief from local-minimum trapping through temperature exchange.

\subsection{Mechanism Analysis via Violation Dynamics}

\begin{figure}[t!]
    \centering
    \includegraphics[width=0.65\linewidth]{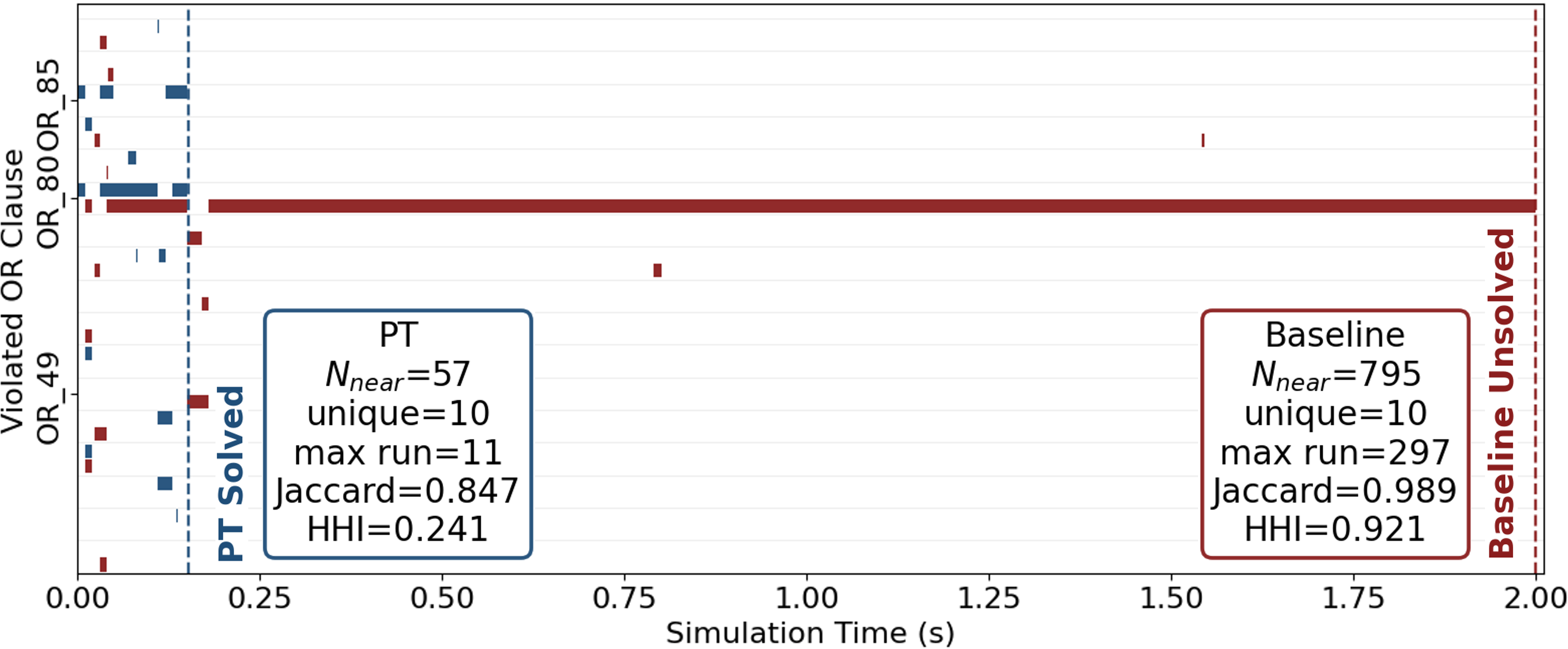}
    \caption{
    Evolution of violated clauses for the hardest benchmark instance, uf20-0192, over time. Horizontal rows correspond to OR clauses; colored segments indicate violation intervals. The parallel baseline is trapped in a local minimum, manifesting as persistent violation patterns 
    ($HHI= 0.921$, $\max\ \text{run}=297$). PT escapes through temperature exchange, shows diverse violation patterns ($HHI=0.241$, $\max\ \text{run}=11$), and reaches a satisfying assignment earlier. }
    \label{fig:local_minima}
\end{figure}

To understand the mechanism behind this performance difference, Figure~\ref{fig:local_minima} examines the runtime dynamics of both solvers on instance uf20-0192, the hardest instance in the benchmark for the parallel baseline. Specifically, Figure~\ref{fig:local_minima} shows violated OR clauses over time.

In the baseline run, the solver repeatedly revisits a small set of violated clauses over extended periods. This manifests as long horizontal segments in the clause visualization.The summary statistics quantify the trapping precisely: 795 near-solution configurations are observed but with only 10 unique violation signatures, a maximum consecutive run of 297 identical configurations, and a high concentration Herfindahl–Hirschman index of $HHI=0.921$. The dynamics are confined to a narrow basin of the energy landscape from which independent solvers at fixed temperature cannot escape.

The PT run exhibits qualitatively different behavior. Violated clauses change frequently, long stationary segments are absent, and the solver reaches a satisfying assignment substantially earlier. The corresponding statistics confirm this: 57 near-solution configurations with 10 unique violation signatures, a maximum consecutive run of only 11, and $HHI=0.241$. These results indicate that PT substantially reduces trapping persistence and promotes transitions across a broader set of low-energy regions.

Taken together, these observations identify the mechanism underlying PT's advantage. Temperature exchange enables replicas to access energy regions that independent fixed-temperature solvers do not reach, thereby reducing the local-minimum trapping that limits baseline performance on hard instances.


\label{Evaluation}

\section{Discussion}
The results show that the benefit of parallel tempering (PT) is highly structured rather than uniform across the benchmark suite. On the majority of instances, both PT and the parallel baseline achieve near-ceiling performance, indicating that temperature exchange provides little added value when fixed-temperature sampling is already sufficient. The advantage of PT emerges primarily on the hard subset of instances, where independent fixed-temperature solvers fail more frequently. This distinction is important because it shows that PT does not merely add redundant parallelism. Instead, it specifically addresses a limitation under independent multi-replica execution.

The runtime analyses clarify the source of this advantage. On hard instances, baseline trajectories repeatedly revisit the same near-solution configurations, indicating confinement to narrow basins of the energy landscape. This behavior is reflected in high concentration values and long consecutive runs of identical violation signatures. PT disrupts this cycle by coupling replicas across temperatures. Higher-temperature dynamics make barrier crossing more likely, while lower-temperature dynamics preserve concentration around promising regions. Through replica exchange, these complementary modes of search are combined within a single solver, producing more diverse exploration and more reliable escape from local minima.

An important practical observation is that this improvement comes with very little additional computational cost. The PT mechanism requires only periodic exchange decisions based on scalar energy values for neighboring replicas. Relative to the cost of spike-based simulation, this overhead is negligible. Moreover, because the exchange logic is external to the neural dynamics, it can be handled by a conventional co-processor in neuromorphic systems without interfering with the operation of the SNN itself. This makes PT not only effective in principle, but also realistic to deploy in hybrid neuromorphic platforms.

More broadly, these findings suggest that simply running multiple independent neural samplers in parallel is not sufficient for robust performance on hard CSP instances. The key limitation is not lack of parallelism, but lack of coordinated escape from structured local-minimum traps. PT provides a principled way to introduce such coordination while preserving the original neural sampling formulation. This makes temperature-coupled neural sampling a compelling extension of existing neuromorphic CSP solvers when reliability on hard instances is the primary objective.
\label{Discussion}

\section{Conclusion}
In this paper, we introduced parallel tempering into an SNN-based neural sampling solver for constraint satisfaction problems. By incorporating temperature control into the neural dynamics and coupling replicas through temperature exchange, the proposed method improves robustness on hard SAT instances beyond what can be achieved by a baseline (under equal computational resources) of independent fixed-temperature samplers.

Across the SATLIB benchmark suite, the benefit of PT is concentrated on the subset of difficult instances where local-minimum trapping is the dominant limitation. Runtime analyses further showed that PT reduces this trapping by enabling access to energy regions that independent fixed-temperature dynamics fail to reach. Together, these results demonstrate that parallel tempering is a practical and effective mechanism for strengthening neural sampling-based CSP solving on hard instances.\label{Conclusion}

\bibliographystyle{ieeetr}
\bibliography{references}

@article{pecevski_nevesim_2014,
AUTHOR={Pecevski, Dejan  and Kappel, David  and Jonke, Zeno },     
TITLE={NEVESIM: event-driven neural simulation framework with a Python interface},      
JOURNAL={Frontiers in Neuroinformatics},     
VOLUME={Volume 8 - 2014},
YEAR={2014},
URL={https://www.frontiersin.org/journals/neuroinformatics/articles/10.3389/fninf.2014.00070},
DOI={10.3389/fninf.2014.00070},
ISSN={1662-5196}}

@article{Hukushima_Exchange_Monte_Carlo_1996,
author = {Hukushima ,Koji and Nemoto ,Koji},
title = {Exchange Monte Carlo Method and  Application to Spin Glass Simulations},
journal = {Journal of the Physical Society of Japan},
volume = {65},
number = {6},
pages = {1604-1608},
year = {1996},
doi = {10.1143/JPSJ.65.1604},
}

@article{Kirkpatrick_Optimization_1983,
author = {S. Kirkpatrick  and C. D. Gelatt  and M. P. Vecchi },
title = {Optimization by Simulated Annealing},
journal = {Science},
volume = {220},
number = {4598},
pages = {671-680},
year = {1983},
doi = {10.1126/science.220.4598.671},}

@article{buesing_neural_2011,
	title = {Neural {Dynamics} as {Sampling}: {A} {Model} for {Stochastic} {Computation} in {Recurrent} {Networks} of {Spiking} {Neurons}},
	volume = {7},
	issn = {1553-7358},
	shorttitle = {Neural {Dynamics} as {Sampling}},
	url = {https://dx.plos.org/10.1371/journal.pcbi.1002211},
	doi = {10.1371/journal.pcbi.1002211},
	language = {en},
	number = {11},
	urldate = {2024-12-27},
	journal = {PLoS Computational Biology},
	author = {Buesing, Lars and Bill, Johannes and Nessler, Bernhard and Maass, Wolfgang},
	editor = {Sporns, Olaf},
	month = nov,
	year = {2011},
	pages = {e1002211},
}

@article{jonke_solving_2016,
	title = {Solving {Constraint} {Satisfaction} {Problems} with {Networks} of {Spiking} {Neurons}},
	volume = {10},
	issn = {1662-453X},
	doi = {10.3389/fnins.2016.00118},
	language = {English},
	urldate = {2024-12-11},
	journal = {Frontiers in Neuroscience},
	author = {Jonke, Zeno and Habenschuss, Stefan and Maass, Wolfgang},
	month = mar,
	year = {2016},
	note = {Publisher: Frontiers},
}

@inproceedings{hoos_satlib_2000,
  author    = {Holger H. Hoos and Thomas St{\"u}tzle},
  title     = {SATLIB: An Online Resource for Research on SAT},
  booktitle = {SAT 2000},
  editor    = {I. P. Gent and H. van Maaren and T. Walsh},
  pages     = {283--292},
  publisher = {IOS Press},
  year      = {2000},
  url       = {http://www.satlib.org}
}

@article{fonseca_guerra_using_2017,
	title = {Using {Stochastic} {Spiking} {Neural} {Networks} on {SpiNNaker} to {Solve} {Constraint} {Satisfaction} {Problems}},
	volume = {11},
	issn = {1662-4548},
	doi = {10.3389/fnins.2017.00714},
	language = {eng},
	journal = {Frontiers in Neuroscience},
	author = {Fonseca Guerra, Gabriel A. and Furber, Steve B.},
	year = {2017},
	pmid = {29311791},
	pmcid = {PMC5742150},
	pages = {714},
}

@inproceedings{binas_spiking_2016,
	title = {Spiking analog {VLSI} neuron assemblies as constraint satisfaction problem solvers},
	url = {https://ieeexplore.ieee.org/document/7538992},
	doi = {10.1109/ISCAS.2016.7538992},
	booktitle = {2016 {IEEE} {International} {Symposium} on {Circuits} and {Systems} ({ISCAS})},
	author = {Binas, Jonathan and Indiveri, Giacomo and Pfeiffer, Michael},
	month = may,
	year = {2016},
	note = {ISSN: 2379-447X},
	pages = {2094--2097},
}

@ARTICLE{Davies_Advancing_2021,
  author={Davies, Mike and Wild, Andreas and Orchard, Garrick and Sandamirskaya, Yulia and Guerra, Gabriel A. Fonseca and Joshi, Prasad and Plank, Philipp and Risbud, Sumedh R.},
  journal={Proceedings of the IEEE}, 
  title={Advancing Neuromorphic Computing With Loihi: A Survey of Results and Outlook}, 
  year={2021},
  volume={109},
  number={5},
  pages={911-934},
  doi={10.1109/JPROC.2021.3067593}}

@book{dechter2003CSP,
  title     = {Constraint Processing},
  author    = {Dechter, Rina},
  year      = {2003},
  publisher = {Morgan Kaufmann},
  address   = {San Francisco, CA}
}

@inproceedings{gupta2006sat,
author = {Gupta, Aarti and Ganai, Malay K. and Wang, Chao},
title = {SAT-Based verification methods and applications in hardware verification},
year = {2006},
isbn = {3540343040},
publisher = {Springer-Verlag},
address = {Berlin, Heidelberg},
url = {https://doi.org/10.1007/11757283_5},
doi = {10.1007/11757283_5},
booktitle = {Proceedings of the 6th International Conference on Formal Methods for the Design of Computer, Communication, and Software Systems},
pages = {108–143},
numpages = {36},
location = {Bertinoro, Italy},
}

@inproceedings{shaw1998,
author = {Shaw, Paul},
title = {Using Constraint Programming and Local Search Methods to Solve Vehicle Routing Problems},
year = {1998},
isbn = {3540652248},
publisher = {Springer-Verlag},
address = {Berlin, Heidelberg},
booktitle = {Proceedings of the 4th International Conference on Principles and Practice of Constraint Programming},
pages = {417–431},
numpages = {15},
}

@book{toth2014,
author = {Toth, Paolo and Vigo, Daniele},
title = {Vehicle Routing},
publisher = {Society for Industrial and Applied Mathematics},
year = {2014},
doi = {10.1137/1.9781611973594},
address = {Philadelphia, PA},
edition   = {2},
}

@InProceedings{cambazard2013,
author="Cambazard, Hadrien
and Mehta, Deepak
and O'Sullivan, Barry
and Simonis, Helmut",
editor="Altmann, J{\"o}rn
and Vanmechelen, Kurt
and Rana, Omer F.",
title="Constraint Programming Based Large Neighbourhood Search for Energy Minimisation in Data Centres",
booktitle="Economics of Grids, Clouds, Systems, and Services",
year="2013",
publisher="Springer International Publishing",
address="Cham",
pages="44--59",
isbn="978-3-319-02414-1"
}

@book{hoos2004,
  author    = {Hoos, Holger H. and St{\"u}tzle, Thomas},
  title     = {Stochastic Local Search: Foundations and Applications},
  publisher = {Elsevier/Morgan Kaufmann},
  year      = {2004}
}

@article{pecevski2011,
	title = {Probabilistic {Inference} in {General} {Graphical} {Models} through {Sampling} in {Stochastic} {Networks} of {Spiking} {Neurons}},
	volume = {7},
	url = {https://doi.org/10.1371/journal.pcbi.1002294},
	doi = {10.1371/journal.pcbi.1002294},
	number = {12},
	journal = {PLOS Computational Biology},
	publisher = {Public Library of Science},
	author = {Pecevski, Dejan and Buesing, Lars and Maass, Wolfgang},
	month = dec,
	year = {2011},
	pages = {1--25},
}

@article{hastings1970,
 ISSN = {00063444, 14643510},
 URL = {http://www.jstor.org/stable/2334940},
 author = {W. K. Hastings},
 journal = {Biometrika},
 number = {1},
 pages = {97--109},
 publisher = {[Oxford University Press, Biometrika Trust]},
 title = {Monte Carlo Sampling Methods Using Markov Chains and Their Applications},
 urldate = {2026-04-09},
 volume = {57},
 year = {1970}
}

\end{document}